\documentstyle[12pt]{article}
\def\beq{\begin{equation}}
\def\eeq{\end{equation}}
\def\bea{\begin{eqnarray}}
\def\eea{\end{eqnarray}}
\begin {document}
\begin{titlepage}
Revised version, September 1994 \hfill DESY 94-066\\
\begin{flushright}
HU Berlin-IEP-94/2 \\
\end{flushright}
\mbox{ }  \hfill hepth@xxx/9403141
\vspace{6ex}
\Large
\begin {center}
\bf{Two and three-point functions in Liouville theory}
\end {center}
\large
\vspace{3ex}
\begin{center}
H. Dorn and H.-J. Otto
\footnote{e-mail: dorn@ifh.de , otto@ifh.de}
\end{center}
\normalsize
\it
\vspace{3ex}
\begin{center}
Humboldt--Universit\"at zu Berlin \\
Institut f\"ur Physik, Theorie der Elementarteilchen \\
Invalidenstra\ss e 110, D-10115 Berlin
\end{center}
\vspace{6 ex }
\rm
\begin{center}
\bf{Abstract}
\end{center}
\vspace{3ex}
Based on our generalization of the Goulian-Li continuation in the power of
the 2D cosmological term we construct the two and three-point correlation
functions for Liouville exponentials with generic real coefficients. As a
strong argument in favour of the procedure we prove the Liouville equation
of motion on the level of three-point functions. The analytical structure
of the correlation functions as well as some of its consequences for string
theory are discussed. This includes a conjecture on the mass shell condition
for excitations of noncritical strings. We also make a comment concerning
the correlation functions of the Liouville field itself.
\end {titlepage}
\newpage
\setcounter{page}{1}
\pagestyle{plain}
\section {Introduction}
Noncritical string theory and more general conformal field theories coupled
to 2D gravity  require the knowledge of Liouville theory which describes
the Weyl degree of freedom of the 2D metrics. In addition, as a highly
nontrivial interacting 2D field theory Liouville theory deserves a lot of
attention for its own. A central problem is the construction of arbitrary
real powers of the exponential of the Liouville field. They are needed as
gravitational dressing factors \cite{David,DK,DO1} for vertex operators in
string theory or conformal models beyond the minimal series. Furthermore,
generic real powers of the Liouville exponential can be used to define
via differentiation the Liouville operator itself. \\

In the canonical approach based on free fields \cite{Gervais1,OW,KN}
(and refs.
therein) a lot of important structure, including the underlying
quantum group \cite{QG} has been investigated. The general exponential under
discussion
is given by an infinite series of operators constructed out of free fields
\cite{GSchn}. However, the calculation of correlation functions based on this
representation is still lacking.\\

We approach the problem of Liouville correlation functions in this paper
along a less rigorous way following the idea of Goulian and Li \cite{GL}.
It implies a continuation from integer values of a certain variable $s$
to arbitrary real values. Applied to minimal conformal models the procedure
reproduces the results of corresponding matrix model calculations. It has
been justified by comparison of its asymptotics with that obtained for
the original functional integral in the quasiclassical limit \cite{AH}.
There is also a way to understand the success of the method within
the canonical treatment \cite{Gervais2}.\\
The original method of \cite{GL} was developed only for those exponentials
which are needed as dressing factors for primary fields in minimal
conformal models, i.e. for special rational values of $s$.
In the derivation explicit use has been made of constraints
induced by the special degenerate kinematics in one-dimensional target space.
In a previous paper \cite{DO2} we have constructed a continuation free of
these restrictions.\\

Using our continuation procedure and starting from the functional integral
representation we present in section 2 a systematic
construction of three and two-point functions of arbitrary powers of
Liouville exponentials. In section 3 we check the Liouville equation for
the three-point function. The uniqueness proof of \cite{AH} is restricted
to exponentials with rational coefficients as they are necessary to dress
minimal models. With a more general validity proof still lacking
we understand the check of the Liouville equation as a very strong argument
in favour of the correctness of our continuation procedure. Section 4 is
devoted to a remark on the two and three-point functions of the Liouville
field itself. Section 5 describes the analytic structure of the correlation
functions derived before. Based on this we comment some related work on
off-shell
critical strings \cite{MP} and formulate a conjecture concerning the
mass shell condition for excitations of noncritical strings.

\section{Three and two-point functions of arbitrary Liouville exponentials}
We define the correlation functions under discussion for $N\geq 3$ by
\beq
G_{N}(z_{1},...,z_{N}\vert \beta_{1},...,\beta_{N})~=~ \langle \prod _{j=1}
^{N}e^{\beta_{j}\phi (z_{j})} \rangle
{}~=~ \int D\phi ~e^{-S_{L}[\phi \vert \hat{g}]} \prod _{j=1}^{N}
e^{\beta_{j} \phi (z_{j})}
\label{1}
\eeq
with
\beq
S_{L}[\phi \vert \hat{g}]~=~\frac{1}{8\pi} \int d^{2}z \sqrt{\hat{g}}
\Big(\hat{g}^{mn}\partial_{m}
\phi \partial_{n}\phi~+~Q\hat{R}(z)\phi(z)~+~\mu^{2}e^{\alpha \phi (z)}\Big)~.
\label{2}
\eeq
$\hat{g}$ is a classical reference metric of the 2D manifold of spherical
topology, $\hat{R}$ the corresponding Ricci scalar. $Q$ parametrizes the
central charge $c_{L}$ of the Liouville theory by
\beq
c_{L}~=~1~+~3Q^{2}~.
\label{3}
\eeq
If the Liouville theory describes the gravitational sector of a conformal
matter theory with central charge $c_{M}$ one has in addition
\beq
c_{L}~+~c_{M}~-~26~=~0~.
\label{4}
\eeq
The factor $\alpha$ in the exponential is fixed to attribute to
$e^{\alpha \phi}$ a conformal $(1,1)$ dimension and the property of being
a ``microscopic" operator, i. e. $\alpha <\frac{Q}{2}$ \cite{Seiberg}
\beq
\frac{1}{2} \alpha _{\pm}(Q-\alpha_{\pm})~=~1~~,
\label{5}
\eeq
\beq
\alpha_{\pm}~=~\frac{Q}{2}~\pm ~\frac{\sqrt{Q^{2}-8}}{2}~,~~~~~~~~~\alpha~=~
\alpha_{-}~.
\label{6}
\eeq
The zero mode integration can be performed explicitly \cite{GL}. For integer
\beq
s_{N}~=~\frac{Q-\sum_{j=1}^{N}\beta_{j}}{\alpha}
\label{7}
\eeq
also the remaining functional integral can be done:
\footnote{Effectively we can handle $\mu^{2},~z_{j},~w_{I}$ as dimensionless
quantities, since in addition to the original dimensional $\mu^{2},~z_{j},~
w_{I}$ further dimensional parameters are involved (RG-scale, scale of the
background $\hat{R}$ and an
integration constant in the Liouville action) which can be used to introduce
suitable quotients. A complete discussion of this point is given in the second
reference of \cite{DO1}.}
\bea
G_{N}(z_{1},...,z_{N} \vert \beta_{1},...,\beta_{N})~=~\frac{\Gamma (-s_{N})}
{\alpha} \Big( \frac{\mu ^{2}}{8\pi} \Big) ^{s_{N}}~\prod _{1\leq i<j\leq N}
\vert z_{i}~-~z_{j} \vert ^{-2\beta_{i} \beta_{j}} \nonumber \\
\cdot \int \prod _{I=1}^{s_{N}} \Big( d^{2}w_{I} \prod _{j=1}^{N}\vert z_{j}~-~
w_{I}\vert ^{-2\alpha \beta_{j}}\Big) \prod _{1\leq I<J\leq s_{N}}
\vert w_{I}~-~w_{J}\vert ^{-2\alpha^{2}}~.
\label{8}
\eea
Due to the nonexistence of a SL(2,C) invariant vacuum one has to be careful
with respect to the usual conformal structure of $N$-point functions.\\

Let us start with the 3-point function.
Fortunately from the explicit representation (\ref{8}) we
can prove for integer $s_{3}$ that the standard structure of the $z_{j}$
dependence is realized:\\
At first the integral in (\ref{8}) is convergent for small $\vert \beta _{i}
\vert $ and large $Q$. Then with allowed substitutions of integration
variables one gets first in this region and later by continuation everywhere
for M\"obius transformations $z_{i}~=~V(u_{i})$
\beq
G_{3}(z_{1},z_{2},z_{3} \vert \beta _{1},\beta_{2},\beta_{3})~=~
\prod _{i=1}^{3} \vert V'(u_{i})\vert ^{-2\Delta_{i}}~G_{3}(u_{1},
u_{2},u_{3}
\vert \beta_{1},\beta_{2},\beta_{3})
\label{9}
\eeq
with
\beq
\Delta_{i}~=~\frac{1}{2} \beta _{i}(Q~-~\beta_{i})~.
\label{10}
\eeq
This yields enough information to establish the usual conformal structure
\bea
G_{3}(z_{1},z_{2},z_{3} \vert \beta _{1},\beta_{2},\beta_{3})&=&
A_{3}(\beta_{1},\beta_{2},\beta_{3})~
\vert z_{1}-z_{2} \vert ^{2(\Delta _{3}-\Delta _{1}-\Delta _{2})}
\nonumber \\
&\cdot &\vert z_{1}-z_{3} \vert ^{2(\Delta _{2}-\Delta _{1}-\Delta _{3})}
\vert z_{2}-z_{3} \vert ^{2(\Delta _{1}-\Delta _{2}-\Delta _{3})}~,
\label{11}
\eea
\beq
A_{3}~=~\lim_{u_{3}\to \infty}\vert u_{3}\vert ^{4\Delta _{3}}G_{3}
(0,1,u_{3}\vert \beta_{1},\beta_{2},\beta_{3})~.
\label{12}
\eeq
With (\ref{8}) this gives $A_{3}$ as
\beq
A_{3}~=~\frac{\Gamma(-s_{3})}{\alpha} \Big ( \frac{\mu ^{2}}{8\pi}\Big )
^{s_{3}}\int \prod _{I=1}^{s_{3}}(d^{2}w_{I}\vert w_{I}\vert ^{-2\alpha \beta _
{1}}\vert 1-w_{I} \vert ^{-2\alpha \beta _{2}})
\prod _{1\leq I<J\leq s_{3}}\vert w_{I}-w_{J} \vert ^{-2\alpha ^{2}}~.
\label{13}
\eeq
Using the Dotsenko-Fateev integrals \cite{DF} this can be written as \cite{DO2}
\beq
A_{3}(\beta _{1},\beta _{2},\beta _{3})~=~
\frac{\Gamma (-s_{3})}{\alpha}~\Gamma (1+s_{3})\Big (\frac{\mu ^{2}~ \Gamma
(1+\frac{\alpha ^{2}}{2})}{8~\Gamma(-\frac{\alpha ^{2}}{2})}\Big ) ^{s_{3}}
\prod _{i=0}^{3}F_{i}
\label{14}
\eeq
with
\footnote{$F_{0}$ in the present paper differs from that in ref. \cite{DO2}
by a factor $\pi (-1)^{s_{3}+1}$. The reason is that, as discussed in
\cite{DO2} in detail, in going from integer $s$ to noninteger $s$
one has to give meaning to the formal quantity $\Gamma (0) \sin (\pi s)$.
Instead of $\Gamma (0) \sin (\pi s) \rightarrow 1$ we use now the more
natural convention $ \sin (\pi s) \Gamma (0) \rightarrow
\pi (-1)^{s+1}$. The sign effect will be crucial in section 3 to ensure
the validity of the Liouville equation.}
\beq
F_{i}~=~\exp \Big (f(\alpha \bar{\beta _{i}},\frac{\alpha ^{2}}{2} \vert
s_{3})-
f(\alpha \beta _{i}, \frac{\alpha ^{2}}{2}\vert s_{3})\Big )~,~~~~~~~~~~
i=1,2,3
\label{15}
\eeq
\beq
\bar{\beta _{i}}~=~\frac{1}{2}(\beta _{j}+\beta _{k}-\beta _{i})~
=~\frac{1}{2}(Q-\alpha s_{3})-\beta _{i}~,
{}~~~~~~(i,j,k)=\mbox{perm}(1,2,3)~,
\label{16}
\eeq
\beq
F_{0}~=~\Big (-\frac{\alpha ^{2}}{2}\Big ) ^{-s_{3}}\frac{1}{\Gamma (1+s_{3})}
\exp \Big ( f(1-\frac{\alpha ^{2}}{2}s_{3},\frac{\alpha ^{2}}{2} \vert s_{3})
-f(1+\frac{\alpha ^{2}}{2},\frac{\alpha ^{2}}{2} \vert s_{3})\Big )~,
\label{17}
\eeq
\beq
f(a,b\vert s)~=~\sum_{j=0}^{s-1} \mbox{log}~ \Gamma (a+bj)~,~~~~~\mbox{integer}
{}~s~.
\label{18}
\eeq
The function $f$ fulfills the relations
\beq
f(a,b\vert s+1)=f(a,b\vert s)+\log \Gamma (a+bs)~,
\label{19}
\eeq
\beq
f(a+1,b\vert s)=f(a,b\vert s)+s\log b +\log \Gamma (\frac{a}{b}+s)-
\log \Gamma (\frac{a}{b})~,
\label{20}
\eeq
\beq
f(a+b,b\vert s)=f(a,b\vert s)+\log \Gamma (a+bs)-\log \Gamma (a)~,
\label{21}
\eeq
\beq
f(a+b(s-1),-b\vert s)=f(a,b\vert s)~.
\label{22}
\eeq
Further functional relations can be found in \cite{DO2}, we present
here only those we need in the discussion of the present paper.
There exists a continuation of $f(a,b\vert s)$ to arbitrary complex $a,b,s$
given by
\bea
f(a,b\vert s)=\int_{0}^{\infty }\frac{dt}{t}\Big(s(a-1)
e^{-t}+b\frac{s(s-1)}{2}
e^{-t}-s\frac{e^{-t}}{1-e^{-t}}
\nonumber\\
+\frac{(1-e^{-tbs})e^{-at}}{(1-e^{-tb})
(1-e^{-t})}\Big).
\label{23}
\eea
It fulfills all the functional relations. Using the integral representation
and the functional relations one can prove \cite{DO2} that exp$(f(a,b\vert s))$
is a meromorphic function. Due to (\ref{22}) it is sufficient to investigate
the case Re $b\geq 0$. Under this circumstance exp$f$ has poles at
\beq
a~=~-bj~-~l~~~~~~~~~~\mbox{(poles)}
\label{24}
\eeq
and zeros at
\beq
a~+~bs~=~-bj~-~l~~~~~~~~~~\mbox{(zeros)}~.
\label{25}
\eeq
In both cases $j$ and $l$ are integers $\geq 0$.
The order of poles and zeros is determined by the number of different
realizations of the r.h.s. of eqs. (\ref{24}) and (\ref{25}), respectively.\\

We now turn to the 2-point function. Taking (\ref{1}) unmodified also for $N=2$
would imply $G_{2}(z_{1},z_{2}\vert \beta _{1},\beta _{2})~=~
G_{3}(z_{1},z_{2},z_{3}\vert \beta _{1},\beta _{2},0)$. The unwanted
$z_{3}$-dependence as usual in conformal theories drops for $\Delta _{1}=
\Delta _{2}$. However, the $z$-independent factor $A_{3}(\beta ,\beta ,
\beta _{3})$ diverges for $\beta _{3} \rightarrow 0$. This can be seen
from
\bea
\prod _{j=1}^{3} F_{j}&=&\prod _{j=1}^{3}\Big ( \exp [f(\alpha \bar{\beta
_{j}} +1,\frac{\alpha ^{2}}{2}\vert s_{3})-f(\alpha \beta _{j}+1,\frac{\alpha
^{2}}{2}\vert s_{3})]~\frac{\Gamma (\frac{2\beta _{j}}{\alpha}+s_{3})\Gamma
(1+\frac{2\bar{\beta _{j}}}{\alpha})}{\Gamma (\frac{2\bar{\beta _{j}}}{
\alpha}+s_{3})\Gamma (1+\frac{2\beta _{j}}{\alpha})}\Big )
\nonumber \\
&\cdot &\frac{8\beta _{1}\beta _{2}\beta _{3}}{(\beta _{1}+\beta _{2}-\beta
_{3})
(\beta _{1}+\beta _{3}-\beta _{2})(\beta _{2}+\beta _{3}-\beta _{1})}~~,
\label{26}
\eea
which is a consequence of (\ref{20}) and (\ref{16}). The reason for this
divergence is the change of the situation with respect to the conformal Killing
vectors (CKV). The 3-punctured sphere has no CKV's while the 2-punctured
sphere has one. The (divergent) volume of the corresponding subgroup
of the M\"obius group SL(2,C) leaving $z_{1}$ and $z_{2}$ fixed is
\beq
V^{(2)}_{CKV}~=~\int \frac{d^{2}w~\vert z_{1}-z_{2}\vert ^{2}}
{\vert z_{1}-w\vert ^{2}~\vert z_{2}-w\vert ^{2}}~~.
\label{27}
\eeq
Having this in mind we define
\beq
G_{2}(z_{1},z_{2}\vert \beta )~=~\langle e^{\beta \phi (z_{1})}e^{\beta \phi
(z_{2})}\rangle ~=~\frac{1}{V^{(2)}_{CKV}}\int D\phi ~e^{-S_{L}[\phi ]}
e^{\beta \phi (z_{1})}e^{\beta \phi (z_{2})}~~.
\label{28}
\eeq
Treating the functional integral in analogy to that for the 3-point function
and
choosing $\int d^{2}w_{1}$ as the cancelled integration one gets
\bea
&G_{2}(z_{1},z_{2}\vert \beta)&=~\frac{\Gamma (-s_{2})}{\alpha}~\Big ( \frac
{\mu ^{2}}{8\pi}\Big )^{s_{2}}\vert z_{1}~-~z_{2}\vert ^{-2\beta ^{2}}~
\frac{\vert z_{1}-w_{1}\vert ^{2}~\vert z_{2}-w_{1}\vert ^{2}}{\vert z_{1}-
z_{2}\vert ^{2}}~~~~~~~~~~~~~~~~~~~~
\label{29} \\
&&\cdot \int \prod _{I=2}^{s_{2}}d^{2}w_{I}\prod _{I=1}^{s_{2}}\Big (
\vert z_{1}-w_{I} \vert ~\vert z_{2}-w_{I}\vert \Big )^{-2\alpha
\beta}\prod _{1\leq I<J\leq s_{2}}~\vert w_{I}-w_{J}\vert ^{-2\alpha ^{2}}.
\nonumber
\eea
This means (note $s_{2}=\frac{Q-2\beta}{\alpha}=1+s_{3}(\beta ,\beta ,\alpha)$)
\beq
G_{2}(z_{1},z_{2}\vert \beta)~=~-\frac{\mu ^{2}}{8\pi
s_{2}}G_{3}(z_{1},z_{2},w_{1}\vert \beta ,\beta ,\alpha)~\vert z_{1}-z_{2}
\vert
^{-2}~\vert z_{1}-w_{1}\vert ^{2}~\vert z_{2}-w_{1}\vert ^{2}~.
\label{30}
\eeq
{}From (\ref{11}) we see that the $w_{1}$ dependence on the r.h.s. cancels.
For this result $\Delta _{1}=\Delta _{2}$ is crucially. Altogether we find
\beq
G_{2}(z_{1},z_{2}\vert \beta )~=~\frac{A_{2}(\beta )}{\vert z_{1}-z_{2}\vert
^{4\Delta}}~~,
\label{31}
\eeq
with
\beq
A_{2}(\beta )~=~-\frac{\mu ^{2}}{8\pi s_{2}}~A_{3}(\beta ,\beta ,\alpha )~~.
\label{32}
\eeq
By the use of (\ref{22}) and (\ref{21}) one can eliminate the function $f$
completely and derive the simple result
\beq
A_{2}(\beta )~=~-~\frac{1}{\alpha \pi s_{2}} \Big (\frac{\mu ^{2}~\Gamma (\frac
{\alpha ^{2}}{2})}{8~\Gamma (1-\frac{\alpha ^{2}}{2})}\Big
)^{s_{2}}~\frac{\Gamma (1-\frac{\alpha ^{2}}{2}s_{2})~\Gamma (1-s_{2})}
{\Gamma (\frac{\alpha ^{2}}{2}s_{2})~\Gamma (s_{2})}~~.
\label{33}
\eeq
Due to (\ref{7}) and (\ref{5}) $s_{2}$ expressed by
$\alpha $ and $\beta $ is
$$s_{2}=1+\frac{2}{\alpha
^{2}}-\frac{2\beta}{\alpha}~.$$
This formula for $A_{2}$ coincides up to an irrelevant
factor $-\frac{\pi \alpha ^{11}}
{32}$ with the expression presented in \cite{GL} for the integrated
2-point function in gravitationally dressed minimal models, i.e. for rational
$s_{2}$. For $A_{2}$ describing the gravitational dressing of the two point
function in minimal models the form (\ref{33}) fits into the ``leg-factor"
structure known for higher correlation functions ($N\geq 3$), \cite{FK}.\\

The extension of the procedure to the one and zero-point function (partition
function) is straightforward. One has to cancel $V^{(1)}_{CKV}$ and $
V^{(0)}_{CKV}$ related to the one-punctured and unpunctured sphere,
respectively. However, in the one-point case the method yields inconsistent
results, the dependence on the fixed unintegrated $w_{1},~w_{2}$ does not
cancel for generic $\beta $.
This is a reflection of the absence of a SL(2,C) invariant vacuum,
which prevents looking at the one-point function as a scalar product of two
physical states. We come back to this point in section 4.
For the partition function $Z$ we get
\bea
Z~=~G_{0}&=&-\frac{\mu ^{6}}{512\pi ^{3}s_{0}(1-s_{0})(2-s_{0})}~A_{3}(\alpha ,
\alpha ,\alpha ) \nonumber \\
&=&-\frac{\mu ^{2}}{8\pi ^{3}(\alpha +\frac{2}{\alpha})}~\Big (\frac
{\mu ^{2}~\Gamma (\frac{\alpha ^{2}}{2})}{8~\Gamma(1-\frac{\alpha
^{2}}{2})}\Big )^{\frac{2}{\alpha ^{2}}}~\frac{\Gamma (-\frac{2}{\alpha ^{2}})}
{\Gamma (\frac{2}{\alpha ^{2}}-1)}.
\label{34}
\eea
After this discussion of $G_{2}$ and $G_{0}$ we want to mention a modified
argumentation, which avoids the use of the $w_{I}$ integral representation (A
similar argument for the string S-matrix elements, i.e. integrated correlation
functions of matter $\otimes$ Liouville has been applied in \cite{GL}). From
the
formal functional integral (\ref{1}) one gets by differentiation with respect
to
$\mu ^{2}$ that $-8\pi \frac{d}{d\mu ^{2}}G_{2}(z_{1},z_{2}\vert \beta )$ has
to
be equal to the once integrated 3-point function $\int d^{2}z_{3}
G_{3}(z_{1},z_{2},z_{3}\vert \beta ,\beta ,\alpha )$. To complete this to an
exact statement we must divide by $V^{(2)}_{CKV}$, i.e.
\beq -8\pi \frac{d}{d\mu ^{2}} G_{2}(z_{1},z_{2}\vert \beta
)~=~\frac{1}{V^{(2)}_{CKV}}\int d^{2}z_{3} ~G_{3}(z_{1},z_{2},z_{3} \vert \beta
, \beta ,\alpha )~~.  \label{35}
\eeq Taking into account $G_{N}\propto \mu ^{2s_{N}}$ one arrives at
(\ref{31}),
(\ref{32}), immediately.\\

Closing this section we want to make a comment on the quasiclassical
approximation $Q\rightarrow \infty$. In this limit one has $\alpha \cdot Q
\rightarrow 2$. To get $Q^{2}$ as an overall factor in front of the action
one should make the usual rescaling $\phi \rightarrow \frac{Q}{2} \phi $.
Hence a sensible quasiclassical limit is
$$\alpha \rightarrow 0,~~~\beta~=~\alpha b,~~~\mu ^{2}~=~\frac{2m^{2}}{
\alpha^{2}},~~~b,m^{2}~ \mbox{fixed.}$$
Under these conditions we find ($C$ Euler constant) from (\ref{33}), (\ref{34})
\bea
A_{2}&=&-\frac{e^{4b-2-2C}}{\pi \alpha ^{3} (2b-1)\sin (\pi (2b-\frac{2}{
\alpha^{2}}))}~\Big (\frac{m^{2}e^{2}}{8} \Big )^{1-2b+\frac{2}{\alpha ^{2}}}
{}~(1+O(\alpha ))~,
\nonumber \\
Z&=&-\frac{e^{-2-2C}}{\pi ^{3}\alpha ^{3} \sin (-\frac{2}{
\alpha^{2}}\pi )}~\Big (\frac{m^{2}e^{2}}{8} \Big )^{1+\frac{2}{\alpha ^{2}}}
{}~(1+O(\alpha ))~.
\label{a1}
\eea
The normalized 2-point function fulfills in the limit
\beq
\frac{G_{2}}{Z}~=~\frac{\pi ^{2}}{2b-1}~\Big (-~\frac{m^{2}}{8} \big )^{-2b}~
\vert z_{1}-z_{2}\vert ^{-4b}~.
\label{a2}
\eeq
Here we have used $\frac{\sin (-\frac{2\pi}{
\alpha ^{2}})}{\sin (\pi (2b-\frac{2}{\alpha ^{2}}))}~\rightarrow ~\exp (\pm
2\pi bi)~=~(-1)^{2b}$, which is valid of course only for
a limit performed a little bit off
the real axis, i.e. $\alpha ^{2}=\vert \alpha ^{2} \vert e^{\pm i\epsilon}$.
Eq. (\ref{a2}) gives the structure presented in ref. \cite{Seiberg}
for the quasiclassical limit in a more explicit form.

\section{Liouville equation of motion}
The Liouville equation in our parametrization is the equation of motion for the
action (\ref{2}) in the limit of flat $\hat{g}$
\beq
\partial ^{2}~\phi ~-~\frac{\alpha \mu ^{2}}{2}~e^{\alpha \phi}~=~0~.
\label{36}
\eeq
As a partial check we want to prove
\beq
\langle \partial ^{2}\phi (z_{1})~e^{\beta _{2}\phi (z_{2})}~
e^{\beta _{3}\phi (z_{3})}\rangle~=~\frac{\alpha \mu ^{2}}{2}\langle
e^{\alpha \phi (z_{1})}~e^{\beta _{2}\phi (z_{2})}~e^{\beta _{3}\phi (z_{3})}
\rangle
\label{37}
\eeq
up to contact terms.\\
The l.h.s. of (\ref{37}) is given by\\
$$4\partial _{z_{1}}\partial _{\bar{z_{1}}}~\lim _{\beta _{1}\rightarrow 0}
\frac{\partial}{\partial \beta _{1}}~G_{3}(z_{1},z_{2},z_{3}\vert \beta _{1},
\beta _{2},\beta _{3})~.$$
Using (\ref{10}), (\ref{11}) the differentiation with respect to $\beta _{1}$
is straightforward. In the generic case $\beta _{2}\neq \beta _{3},~
\beta _{j}\neq 0,~~j=2,3$ one has $A_{3}(0,\beta _{2},\beta _{3})=0$ as
can be seen from eqs. (\ref{14}), (\ref{17}), (\ref{26}). Therefore,
the contribution
of terms with logarithms log$\vert z_{i}-z_{j}\vert $ generated by the
$\beta _{1}$ dependence of $\Delta _{1}$ drops out in the limit $\beta _{1}
\rightarrow 0$, i. e.
\footnote{The $\beta _{1}$ derivative of $A_{3}$ is of course a total one
refering also to the implicit $\beta _{1}$ dependence in $\bar{\beta _{j}}$
and $s_{3}$.}
\beq
\lim _{\beta_{1}\rightarrow 0}\frac{\partial}{\partial \beta _{1}}
G_{3}(~z_{j}~\vert ~\beta _{j}~)~=~
\frac{\frac{\partial}{\partial \beta _{1}}
A_{3}(\beta _{1},\beta _{2},\beta _{3})\vert _{\beta _{1}=0}}
{\vert z_{2}-z_{1}\vert ^{2(\Delta_{2}-\Delta_{3})}
\vert z_{1}-z_{3}\vert ^{2(\Delta_{3}-\Delta_{2})}
\vert z_{2}-z_{3}\vert ^{2(\Delta_{3}+\Delta_{2})} }~.
\label{38}
\eeq
After differentiation with respect to $z_{1}$ this yields
\beq
\langle \partial^{2}\phi (z_{1})e^{\beta _{2}\phi (z_{2})}
e^{\beta _{3}\phi (z_{3})}\rangle ~=~
\frac{4(\Delta _{2}-\Delta _{3})^{2}
\frac{\partial}{\partial \beta _{1}}
A_{3}(\beta _{1},\beta _{2},\beta _{3})
\vert _{\beta _{1}=0}}
{\vert z_{2}-z_{1}\vert ^{2(1+\Delta_{2}-\Delta_{3})}
\vert z_{1}-z_{3}\vert ^{2(1+\Delta_{3}-\Delta_{2})}
\vert z_{2}-z_{3}\vert ^{2(\Delta_{3}+\Delta_{2}-1)} }~
\label{39}
\eeq
up to contact terms.\\
Since the r.h.s. of (\ref{37}) is a special case of the 3-point function,
eq. (\ref{37}) is valid iff
\beq
4(\Delta _{2}-\Delta _{3})^{2}
\frac{\partial}{\partial \beta _{1}}
A_{3}(\beta _{1},\beta _{2},\beta _{3})
\vert _{\beta _{1}=0}~=~\frac{\alpha \mu ^{2}}{2}~A_{3}(\alpha ,\beta _{2},
\beta _{3})~.
\label{40}
\eeq
To prove (\ref{40}) one first observes that for general $\beta _{2},
{}~\beta _{3}$ but $\beta _{1}=0$
all factors in $A_{3}$ except $F_{1}$ are regular
and different from zero. With (\ref{15}), (\ref{20}) we get ($s=s_{3}(0,\beta
_{2},\beta _{3})=\frac{Q-\beta _{2}-\beta _{3}}{\alpha}$)
\bea
\Big (F_{1}\Big )_{\beta _{1}=0}&=&0
\label{41} \\
\Big (\frac{\partial}{\partial \beta _{1}}F_{1}\Big )
_{\beta _{1}=0}&=&\alpha \Big (\frac{\alpha ^{2}}
{2}\Big )^{s-1}~\Gamma (s)~\exp \Big ( f(\frac{\alpha}{2}(\beta _{2}+\beta _{3}
),\frac{\alpha ^{2}}{2}\vert s)~-~f(1,\frac{\alpha ^{2}}{2}\vert s)
\Big )~.
\nonumber
\eea
Therefore, to calculate $\frac{\partial A_{3}}{\partial \beta _{1}}$ for $\beta
_{1}\rightarrow 0$ we need only the derivative of $F_{1}$ \beq
\Big (\frac{\partial A_{3}}{\partial \beta _{1}}\Big )_{\beta _{1}=0}~=~
\frac{\Gamma
(-s)}{\alpha}\Gamma (1+s)\Big (\frac{\mu^{2}~\Gamma (1+\frac{\alpha
^{2}}{2})}{2~\Gamma (-\frac{\alpha^{2}}{2})}\Big )^{s}
F_{0}F_{2}F_{3}\frac{\partial}{\partial \beta _{1}}F_{1}\vert _{\beta _{1}=0}~.
\label{42}
\eeq
Having reached this level, the proof of eq. (\ref{40}) is completed by
a repeated application of the functional relations (\ref{19}), (\ref{20}),
(\ref{21}).
\section{Remark on the three and two-point function of the Liouville field}
As mentioned in the introduction and practised at an intermediate stage in the
previous section already, we relate the Liouville operator $\phi (z)$ to
$\partial _{\beta}e^{\beta \phi (z)}~=~\phi e^{\beta \phi (z)}$.
For reasons becoming clear in a moment
we still do not specify the value of $\beta $ after differentiation. We
only require to treat all $\beta _{j}$ on an equal footing.
{}From (\ref{11}), (\ref{10}), (\ref{6}) this yields
\bea
\partial _{\beta_{1}}\partial _{\beta_{2}}\partial _{\beta_{3}}G_{3}(
{}~z_{j}~\vert ~\beta _{j}~)&=&
\vert z_{1}-z_{2} \vert ^{2(\Delta _{3}-\Delta _{1}-\Delta _{2})}
\vert z_{1}-z_{3} \vert ^{2(\Delta _{2}-\Delta _{1}-\Delta _{3})}
\vert z_{2}-z_{3} \vert ^{2(\Delta _{1}-\Delta _{2}-\Delta _{3})}
\nonumber \\
\cdot &\Big (&\partial _{\beta_{1}}\partial _{\beta_{2}}\partial _{\beta_{3}}
A_{3}~
-~\sqrt {Q^{2}-8}~\partial _{\beta _{1}}\partial _{\beta _{2}}A_{3}(l_{12}~+~
l_{13}~+~l_{23})\nonumber \\
&+&(Q^{2}-8)~\partial _{\beta _{1}}A_{3}~(L_{13}~L_{12}~+~L_{23}~L_{12}~+~
L_{23}~L_{13})\hfill \nonumber \\
&+&(Q^{2}-8)^{\frac{3}{2}}A_{3}~L_{12}~L_{13}~L_{23}~\Big )~, \nonumber \\
\label{43}
\eea
with $l_{ij}~=~\mbox{log}\vert z_{i}-z_{j}\vert,~~~L_{ij}~=~
l_{ij}~-~l_{ki}~-~l_{kj},~~(i,j,k)~=~\mbox{perm}(1,2,3)$.\\

Use has been made of the equality of derivatives with respect to different
$\beta _{j}$ if after differentiation a symmetric point in $\beta $-space
is choosen.\\
The natural choice $\beta _{j}=0$ removes the power-like $z$-dependence
in (\ref{43}), but as can be seen from eqs. (\ref{17}) and
(\ref{26}) $A_{3}$ is singular at this point: The value of $F_{1}F_{2}F_{3}$
depends on how the origin in $\beta $-space is approached. In addition
$F_{0}$ has a pole (note $s_{3}\rightarrow \frac{Q}{\alpha}=1+\frac{2}
{\alpha ^{2}}$, (\ref{24}), (\ref{25})).\\

Our functional integral yields the correlation
function of Liouville exponentials directly, there is no interpretation as
a vacuum expectation value with respect to a SL(2,C) invariant vacuum
\cite{Seiberg}. Therefore, operator insertions are well-defined only in the
presence of at least two
Liouville exponentials playing the role of spectators. This
concept worked perfectly in the previous section where we constructed
$\langle \phi (z_{1})e^{\beta _{2}\phi (z_{2})}e^{\beta _{3}\phi (z_{3})}
\rangle $. To get in the same sense 2 and 3-fold insertions of $\phi $
one has to start with the 4 and 5-point functions of exponentials.
Unfortunately,
these higher correlation functions are not available up to now.\\
There is still another possibility by seeking for an alternative choice
of the $\beta $-value realized after differentiation. The choice $\beta _{j}=
\alpha $ leads to regular $A_{3}$. If one interpretes the remaining
power-like $z$-dependent factor ($\Delta _{j}=1$) as the density of the
M\"obius volume, which has to be cancelled since the $\phi $-insertions
no longer correspond to punctures, one arrives at an expression
for $\langle \phi \phi \phi \rangle $ which is polynomial in $l_{ij}$
up to the third order. In a similar way one can produce a two-point
function containing constant, linear and quadratic terms in $l_{12}$.
However, we have no deeper understanding of such a procedure motivated
purely by technical reasons.
\section{Pole-zero spectrum of the correlation functions}
The spectrum of poles and zeros of the 3-point function and two special
degenerate cases of 4-point functions as well as related problems for the
interpretation of non-critical  strings have been discussed in \cite{DO2},
\cite{DO3}. We add in this section observations concerning the 3-point
and 2-point function which are relevant in connection with some
recent work on off shell critical strings \cite{MP} and which shed some
light on the question of mass shell conditions for non-critical strings.\\

For applications to noncritical strings we are interested in the case
Re $(\alpha ^{2})>0$. This of course is realized for $c_{M}<1$ but higher
dimensional target space $D>1$ made possible by the presence of a linear
dilaton background \cite{DO3} i.e.
\beq
c_{M}=D-3P^{2}.
\label{P}
\eeq
It is even valid
for $1\leq c_{M}<13$ if (\ref{6}) is taken seriously also in
between $1\leq c_{M}\leq 25$, since then with (\ref{3}) and (\ref{4})
we have
\beq
\alpha ^{2}~=~\frac{13~-~c_{M}~-~\sqrt{(25-c_{M})(1-c_{M})}}
{6}~.
\label{44}
\eeq
On the other side in ref. \cite{MP} off shell critical strings are constructed
for $c_{M}=26$ by enforcing the otherwise violated condition of
conformal (1,1) dimension by the dressing with suitable Liouville
exponentials. Clearly, for this application one needs $\alpha ^{2}<0$.\\
In the first situation Re $(\alpha ^{2})>0$ eqs. (\ref{24}) and (\ref{25}) lead
to
the following pole-zero pattern of $\prod _{j=1}^{3}F_{j}$ \cite{DO2}
\newpage
\bea
\alpha ~\bar{\beta _{j}}&=&\frac{\alpha
^{2}}{2}~k_{j}~+~l_{j}~~~~(\mbox{poles})
\nonumber \\
\alpha ~\beta _{j}&=&\frac{\alpha ^{2}}{2}~k_{j}~+~l_{j}~~~~(\mbox{zeros})
\nonumber \\
\mbox{Re}~(\alpha^{2})~>0,~~~~~\mbox{integer}~k_{j},~l_{j},&&\mbox{both}~\leq
0~~~
\mbox{or both}~>0~.
\label{45}
\eea
While the position of zeros depends on the value of single $\beta _{j}$,
the pole position is given by a combination out of all $\beta _{j}$ involved.
Only in applications to dressings of minimal models also the pole position
factorizes (leg poles).\\
In the second situation Re $\alpha ^{2}<0$, since (\ref{24}), (\ref{25})
require Re $b>0$, one must first use the functional relation (\ref{22})
to get
\beq
F_{j}~=~\exp \Big (f(1-\alpha \beta _{j},-\frac{\alpha ^{2}}{2}
\vert s_{3}) ~-~f(1-\alpha \bar{\beta _{j}},-\frac{\alpha ^{2}}{2}
\vert s_{3})\Big )~.
\label{46}
\eeq
Now (\ref{24}), (\ref{25}) are applicable again. Up to a trivial shift
$\beta _{j}$ and $\bar{\beta_{j}}$ change their role
\bea
\alpha \beta _{j}~-~\frac{\alpha ^{2}}{2}&=&-~\frac{\alpha ^{2}}{2}~k_{j}~+
{}~l_{j}~~~~(\mbox{poles})
\nonumber \\
\alpha ~\bar{\beta _{j}}~-~\frac{\alpha ^{2}}{2}&=&-~\frac{\alpha ^{2}}{2}~
k_{j}~+~l_{j}~~~~(\mbox{zeros})
\nonumber \\
\mbox{Re}~(\alpha^{2})~<~0,~~~~~\mbox{integer}~k_{j},~l_{j},&&\mbox{both}~\leq
0~~~
\mbox{or both}~>0~.
\label{47}
\eea
Now the position of poles of $\prod _{j=1}^{3}F_{j}$ is determined by the
single
$\beta _{j}$.\\

The remaining factors in $A_{3}$ depend on the $\beta _{j}$
via $s_{3}$ only. For their combined pole-zero spectrum arising from the
$\Gamma$-functions and $F_{0}$ one finds for Re $\alpha ^{2} >0$ no zeros
but poles at
\bea
\frac{\alpha}{2}\sum _{i=1}^{3}\beta _{i}~-~\frac{\alpha ^{2}}{2}~-&1=&
\frac{\alpha ^{2}}{2}k~+~l~~~~(\mbox{poles})\nonumber \\
\mbox{Re}~(\alpha^{2})~>0,~~~~~\mbox{integer}~k,~l,&&\mbox{both}~\leq 0~~~
\mbox{or both}~>0~.
\label{b1}
\eea
In the other case Re $\alpha ^{2}<0$ one has instead
\bea
\frac{\alpha}{2}\sum _{i=1}^{3}\beta _{i}&=
&\frac{\alpha^{2}}{2}(1-j)~-~1~~~~(\mbox{poles})\nonumber \\
\frac{\alpha}{2}\sum _{i=1}^{3}\beta _{i}&=
&\frac{\alpha^{2}}{2}(k+2)~-~l~~~~(\mbox{zeros})\nonumber \\
\mbox{or}~~~~~
\frac{\alpha}{2}\sum _{i=1}^{3}\beta _{i}&=
&\frac{\alpha^{2}}{2}(1-k)~+~l+2~~~~(\mbox{zeros})\nonumber \\
\mbox{Re}~(\alpha^{2})~<0,~~~&&\mbox{integer}~k,~j,~l \geq 0~.
\label{b2}
\eea
Altogether we find a drastic
change in the analytic structure with respect to the $\beta _{j}$
in going from Re $\alpha ^{2} >0$ to Re $\alpha ^{2}<0$.\\

Let us turn to the 2-point function. From (\ref{33}) we obtain immediately
for arbitrary $\alpha ^{2}$
\bea
\alpha \beta &=&\frac{\alpha ^{2}}{2}~-~l~~\mbox{or}~~~
\alpha \beta~=~1~-~l~\frac{\alpha ^{2}}{2}~,~~\mbox{integer}~l~\geq ~0~~
\mbox{(poles)} \label{48} \\
\alpha \beta &=&\frac{\alpha ^{2}}{2}~+~j~~
\mbox{or}~~\alpha \beta~=~1~+~j~\frac{\alpha ^{2}}{2}~
{}~\mbox{or}~~\beta ~=~\frac{Q}{2}
,~~\mbox{integer}~j~
\geq ~2~~\mbox{(zeros).}
\nonumber
\eea

{}From this pole-zero pattern we can derive an interesting conjecture
concerning
the mass shell condition for noncritical strings. For instance the
coefficient $\beta $ in a gravitationally dressed vertex operator for tachyons
\cite{David,DK,DO1}
$$e^{ik_{\mu}X^{\mu}(z)}~e^{\beta \phi (z)}$$
is related to $k_{\mu}$ by the requirement of total conformal dimension (1,1),
i.e.
$$\frac{1}{2}\beta (Q-\beta )~+~\frac{k (k-P)}{2}~=~1~,$$
or equivalently ((\ref{3}), (\ref{4}), (\ref{P}))
\beq
(\beta - \frac{Q}{2})^{2}~-~(k-\frac{P}{2})^{2}~=~\frac{1-D}{12}~.
\label{49}
\eeq
In contrast to the critical string, where the demand of dimension (1,1)
delivers the mass shell condition $\frac{k(k-P)}{2}=1$, eq. (\ref{49}) implies
no restriction for the target space momentum. \\

A condition on $k_{\mu}$
can arise only due to an additional restriction on the allowed values
of $\beta $.
The $\beta $ dependent factor $A_{2}$ discussed above appears as the dressing
factor in the 2-point S-matrix element for the tachyon excitation of the
string. From the point of view of field theory in target space this object is
an inverse propagator. Hence it should vanish as soon as the tachyon momentum
approaches its mass shell.
\footnote{We assume that this standard reasoning makes sense also in the
presence of a linear dilaton background $P\neq 0$.}
For generic $\beta $ the dressing factor $A_{2}(
\beta )$ is different from zero and it is natural to associate its zeros
with the mass shell. For $c_{M}<1$ i. e. $0<\alpha ^{2}<2$ the spectrum of
zeros
is unbounded from above. The lowest zero is $\beta =\frac{Q}{2}$.
The resulting spectrum for the mass of the gravitational dressed tachyon,
i.e. $m_{T}^{2}=\frac{1-D}{12}-(\beta -\frac{Q}{2})^{2}$, is not bounded
from below. However, since all zeros, except that at $\beta =\frac{Q}{2}$,
obey $\beta >\frac{Q}{2}$ they correspond to operators $e^{\beta \phi}$
describing states with wave functions in mini-superspace approximation
$\propto e^{(\beta -\frac{Q}{2})\phi}$. These states are not normalizable
in the infrared $\phi \rightarrow +\infty $ and have to be excluded \cite
{Seiberg}. On the other hand $\beta ~=~\frac{Q}{2}$ sits just on the border
to the ``microscopic" states describing local insertions with wave functions
peaked
in the ultraviolet and ``macroscopic" states with imaginary exponents.
\footnote{More correctly $\beta ~=~Q/2$ corresponds to the puncture operator
which requires some additional care.}\\
$\beta ~=~Q/2$ then leads to
\beq
(k-\frac{P}{2})^{2}~=~\frac{D-1}{12}~.
\label{50}
\eeq
The generalization to higher string excitations is straightforward.
For instance in the graviton case an additional term $+1$ on the l.h.s.
of (\ref{49}) leads to $(k-\frac{P}{2})^{2}=\frac{D-25}{12}$.\\

We stress that in contrast to discussions based on a $(D+1)$-dimensional
point of view (Liouville field as additional coordinate) on the
l.h.s. of (\ref{50}) we have the $D$-dimensional $k^{2}$ only. Although the
$(D+1)$-dimensional point of view is quite natural in the language of
generalized $\sigma $-model actions \cite{Inder}, for S-matrix elements,
which require the notion of asymptotic states, the situation is more involved.
The $(D+1)$-dimensional concept works perfectly well for vanishing
Liouville mass $\mu $ \cite{FK}. However, for $\mu \neq 0$ the singularity
structure of the special cases of 4-point functions we investigated in
\cite{DO2,DO3} seems to be a serious obstruction for a $(D+1)$-dimensional
interpretation of (\ref{49}).
These special cases ($\beta _{4}~=~0$ or $\alpha$) turned out to be
unsatisfactory from the $D$-dimensional
point of view too. But since both do not fit into the mass shell condition
conjectured in the present paper the issue remains open and requires
more work on the 4-point function.\\

Closing this section we add an observation on the corresponding 3-point string
S-matrix element. Due to (\ref{45}) for $\beta _{1}, \beta _{2}, \beta _{3}$
taking arbitrary values out of the zero table of (\ref{48}), corresponding
to candidates for mass shell values, the function $A_{3}$ has
three zeros, each for every $F_{j}$. A nonvanishing
$A_{3}$ is possible only if there would be in addition three coinciding
poles according to (\ref{45}). By explicit inspection of all possible
cases one can exclude this situation. This property of $A_{3}$
is crucial for further work on the factorization of higher
string S-matrix elements.

\section{Concluding remarks}
With this paper we contributed to the construction of correlation functions in
Liouville theory. This construction is a long standing problem relevant
for various aspects of string theory and general conformal field theory.
We were able to calculate the 2 and 3-point functions of Liouville exponentials
of arbitrary real power. The method of continuation in the parameter $s$
passed a very crucial test. The Liouville equation of motion is fulfilled,
hence we are sure that the derived correlation functions indeed reflect
some essential features of quantized Liouville theory.
What concerns applications to noncritical string theory an interesting
conjecture on mass shell conditions emerged. Keeping the standard
picture that 2-point S-matrix elements vanish on shell, we
related the on shell condition to the spectrum of zeros of the 2-point
function of Liouville exponentials.
The further check of both the $s$-continuation itself as well as the spectrum
conjecture requires the knowledge of the higher ($N\geq 4$) correlation
functions. Unfortunately, at present the necessary integral formulas are
not available. However, judging this unsatisfactory state of affairs
in the $s$-continuation approach one should take into account that
the canonical operator approach still faces problems with the explicit
calculation of the 2 and 3-point function.
\\[10mm]
\noindent
{\bf Acknowledgement}\\
We thank J. Schnittger and G. Weigt for useful discussions.

\end{document}